\documentclass[preprint,preprintnumbers,amsmath,amssymb,superscriptaddress]{revtex4}

\usepackage{graphicx}
\usepackage{dcolumn}
\usepackage{bm}

\begin{document}

Published version: \textbf{Y. Xia \textit{et.al.}, Nature Physics \textbf{5}, 398 (2009)}.

"Topological Insulators: Large-bandgap family found" Nat.Physics Cover (June, 2009).

"Topological Insulators: The Next Generation" Nature Physics \textbf{5}, 378 (2009).

[http://dx.doi.org/10.1038/nphys1294]

\preprint{}

\title{Discovery (theoretical prediction and experimental observation) of a large-gap topological-insulator class with spin-polarized single-Dirac-cone on the surface}

\author{Y. Xia}

\affiliation{Department of Physics, Princeton University, Princeton, NJ 08544, USA}
\affiliation{Princeton Center for Complex Materials, Princeton University, Princeton, NJ 08544, USA}
\author{D. Qian}
\affiliation{Department of Physics, Princeton
University, Princeton, NJ 08544, USA}
\affiliation{Department of Physics, Shanghai Jiao Tong University, Shanghai 200030, China}
\author{D. Hsieh}
\affiliation{Department of Physics,
Princeton University, Princeton, NJ 08544, USA}
\affiliation{Princeton Center for Complex Materials, Princeton University, Princeton, NJ 08544, USA}
\author{L. Wray}
\affiliation{Department of Physics, Princeton University, Princeton, NJ 08544, USA}
\author{A. Pal}
\affiliation{Department of Physics, Princeton University, Princeton, NJ 08544, USA}
\author{H. Lin}
\affiliation{Department of Physics, Northeastern University, Boston, MA 02115, USA}
\author{A. Bansil}
\affiliation{Department of Physics, Northeastern University, Boston, MA 02115, USA}
\author{D. Grauer}
\affiliation{Department of Chemistry, Princeton University, Princeton, NJ 08544, USA}
\author{Y. S. Hor}
\affiliation{Department of Chemistry, Princeton University, Princeton, NJ 08544, USA}
\author{R. J. Cava}
\affiliation{Department of Chemistry, Princeton University, Princeton, NJ 08544, USA}
\author{M. Z. Hasan}
\affiliation{Department of Physics, Princeton University, Princeton, NJ 08544, USA}
\affiliation{Princeton Center for Complex Materials, Princeton University, Princeton, NJ 08544, USA}

\date{Submitted for publication in December 2008}

\maketitle

\textbf{Recent theories and experiments have suggested that strong spin--orbit coupling effects in certain band insulators can give rise to a new phase of quantum matter, the so-called topological insulator, which can show macroscopic entanglement effects \cite{1,2,3,4,5,6,7}. Such systems feature two-dimensional surface states whose electrodynamic properties are described not by the conventional Maxwell equations but rather by an attached axion field, originally proposed to describe strongly interacting particles \cite{8,9,10,11,12,13,14,15}. It has been proposed that a topological insulator \cite{2} with a single spin-textured Dirac cone interfaced with a superconductor can form the most elementary unit for performing fault-tolerant quantum computation \cite{14}. Here we present an \textit{angle-resolved photoemission spectroscopy} study and \textit{first-principle theoretical calculation-predictions} that reveal the first observation of such a topological state of matter featuring a single-surface-Dirac-cone realized in the naturally occurring Bi$_2$Se$_3$ class of materials. Our results, supported by our theoretical predictions and calculations, demonstrate that undoped compound of this class of materials can serve as the parent matrix compound for the long-sought topological device where in-plane surface carrier transport would have a purely quantum topological origin. Our study further suggests that the undoped compound reached via \textit{n}-to-\textit{p} doping should show topological transport phenomena even at room temperature.}

\bigskip
\bigskip
\bigskip
\bigskip
\bigskip
\bigskip
\bigskip
\bigskip
\bigskip
\bigskip
\bigskip
\bigskip
\bigskip
\bigskip
\bigskip
\bigskip
Note added: \textit{The counter-doping method to reach the fully undoped compound - the bulk-insulator state, proposed in this paper, has been achieved in a recent work in Nature:} [http://dx.doi.org/10.1038/nature08234 (2009)]

\newpage

It has been experimentally shown that spin-orbit coupling can lead to new phases of quantum matter with highly nontrivial collective quantum effects \cite{4,5,6}. Two such phases are the quantum spin Hall insulator \cite{4} and the strong topological insulator \cite{5,6,7} both realized in the vicinity of a Dirac point but yet quite distinct from graphene \cite{16}. The strong-topological-insulator phase contains surface states (SSs) with novel electromagnetic properties \cite{7,8,9,10,11,12,13,14,15}. It is currently believed that the Bi$_{1-x}$Sb$_{x}$ insulating alloys realize the only known topological-insulator phase in the vicinity of a three-dimensional Dirac point \cite{5}, which can
in principle be used to study topological electromagnetic and
interface superconducting properties \cite{8,9,10,14}. However, a particular challenge for the topological-insulator Bi$_{1-x}$Sb$_{x}$ system is that the bulk gap is small and the material contains alloying disorder, which makes it difficult to gate for the manipulation and control of charge carriers to realize a device. The topological insulator Bi$_{1-x}$Sb$_{x}$ features five surface bands, of which only one carries the topological quantum number \cite{6}. Therefore, there is an extensive world-wide search for topological phases in stoichiometric materials with no alloying disorder, with a larger gap and with fewer yet still odd-numbered SSs that may work as a matrix material to observe a variety of topological quantum phenomena.

The topological-insulator character of Bi$_{1-x}$Sb$_x$ \cite{5,6} led us to investigate the alternative Bi-based compounds Bi$_2$X$_3$ (X=Se, Te). The undoped Bi$_{2}$Se$_{3}$ is a semiconductor that belongs to the class of thermoelectric materials Bi$_2$X$_3$ with a rhombohedral crystal structure (space group D$^5$$_{3d}$ (R$\bar{3}$m); refs 17,18). The unit cell contains
five atoms, with quintuple layers ordered in the Se(1)-Bi-Se(2)-Bi-Se(1) sequence. Electrical measurements report that, although the bulk of the material is a moderately large-gap semiconductor, its charge transport properties can vary significantly depending on the sample preparation conditions \cite{19}, with a strong tendency to be n-type \cite{20,21} owing to atomic vacancies or excess selenium. An
intrinsic bandgap of approximately 0.35 eV is typically measured in experiments \cite{22,23}, whereas theoretical calculations estimate the gap to be in the range of 0.24-0.3 eV (refs 20, 24).

\begin{figure*}[h!]
\includegraphics[scale=.7]{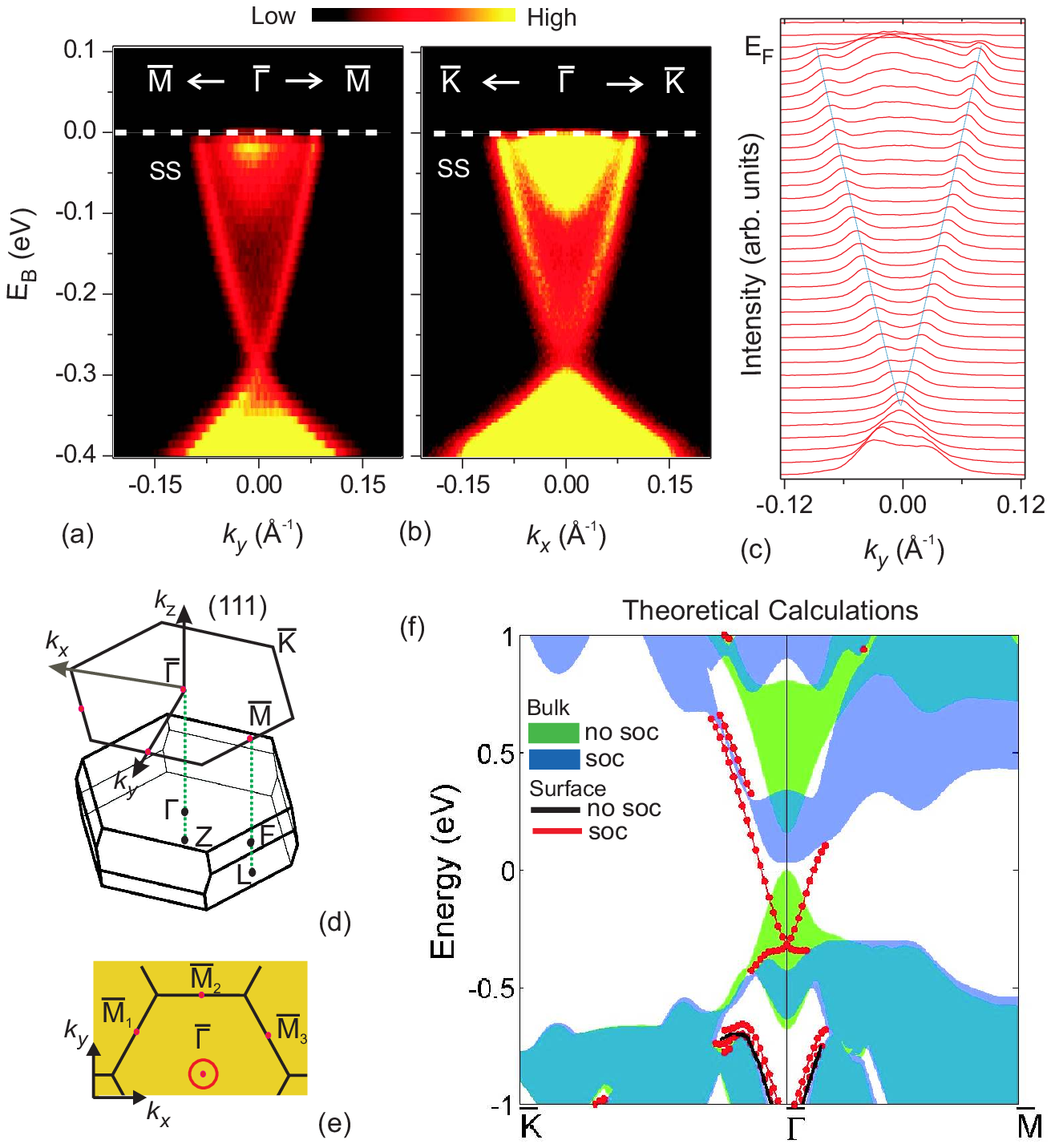}
\caption{\label{fig:charging} \textbf{Theoretical calculations and experimental results : Strong spin-orbit interaction gives rise to a single-surface-state-Dirac-cone on the surface of the topological insulator.}}
\end{figure*}

\begin{figure*}[h!]
\includegraphics[scale=0.7]{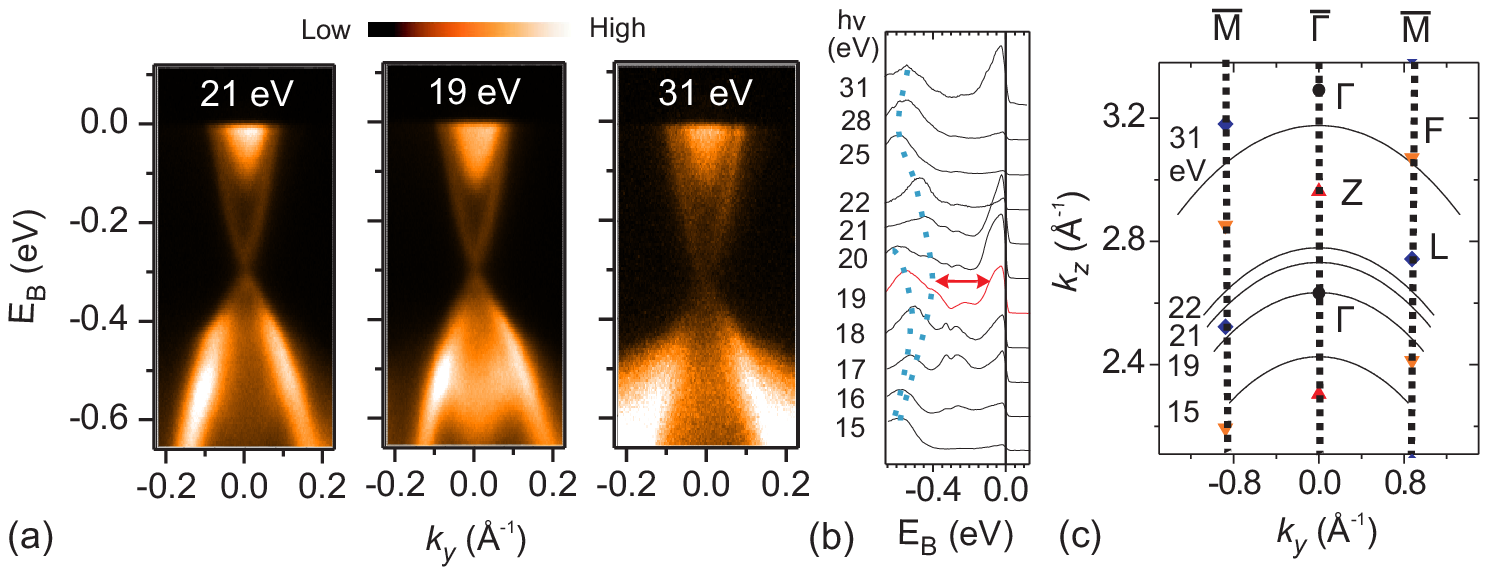}
\caption{\label{fig:charging} \textbf{Transverse-momentum $k_z$ dependence of topological Dirac bands near $\bar{\Gamma}$.}}
\end{figure*}

\begin{figure*}[h!]
\includegraphics[scale=0.70]{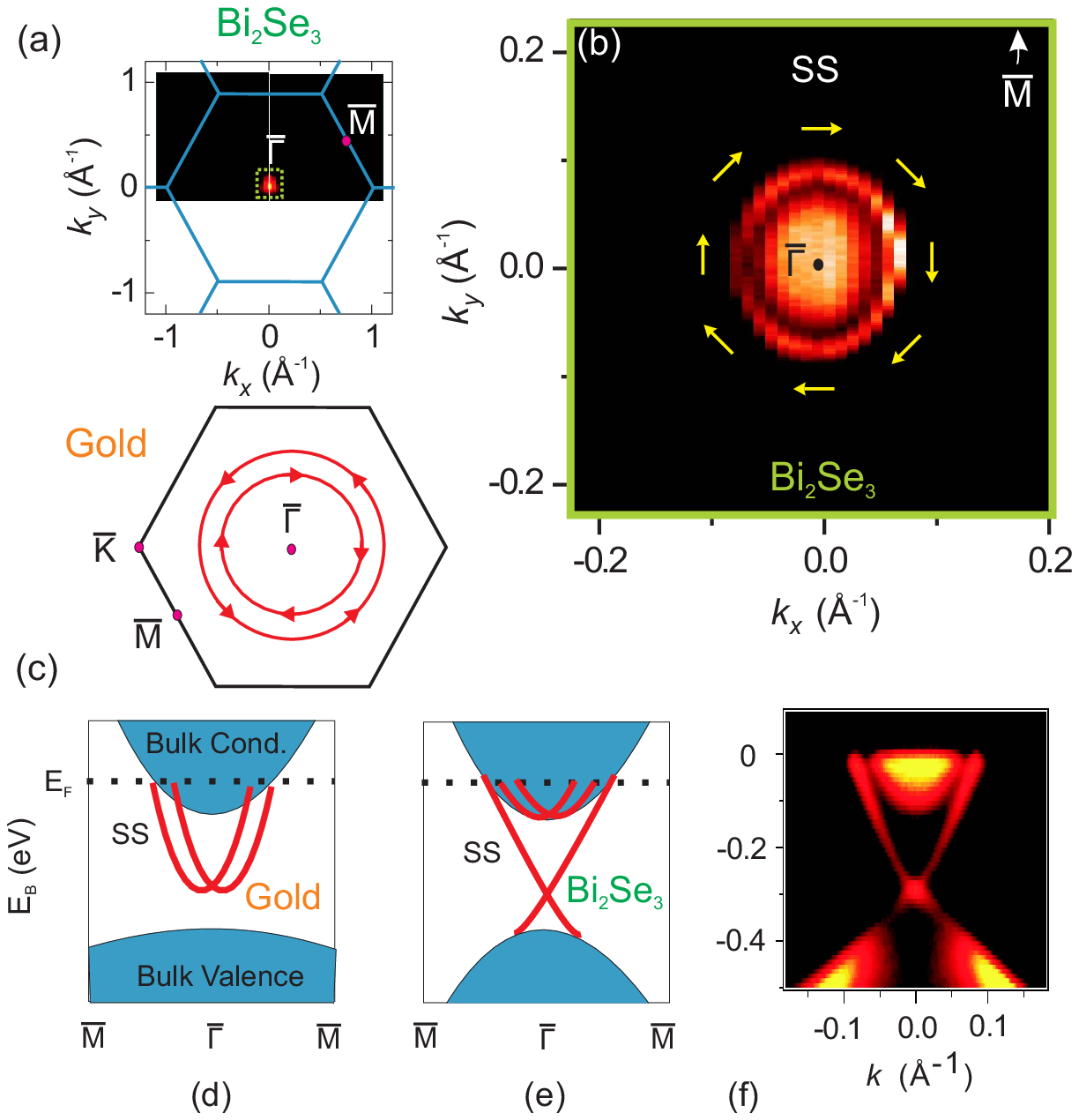}
\caption{\label{fig:charging} \textbf{The topology of the surface Dirac cone Fermi surface and the associated Berry's phase.}}
\end{figure*}

It has been shown that spin-orbit coupling can lead to
topological effects in materials that determine their spin Hall
transport behaviors \cite{4,5,6,7}. Topological quantum properties are directly probed from the nature of the electronic states on the surface by studying the way surface bands connect the material's bulk valence and conduction bands in momentum space \cite{5,6,7}. The surface electron behavior is intimately tied to the number of bulk band inversions that exist in the band structure of a material \cite{7}. The origin of topological Z$_2$ order in Bi$_{1-x}$Sb$_{x}$ is bulk-band inversions
at three equivalent L-points \cite{5,7} whereas in Bi$_{2}$Se$_{3}$ only one band is expected to be inverted, making it similar to the case in the two-dimensional quantum spin Hall insulator phase. Therefore, a much simpler surface spectrum is naturally expected in Bi$_{2}$Se$_{3}$.
All previous experimental studies of Bi$_{2}$Se$_{3}$ have focused on the material's bulk properties; nothing is known about its SSs. It is this key experimental information that we provide here that, for the first time, enables us to determine its topological quantum class.

The bulk crystal symmetry of Bi$_{2}$Se$_{3}$ fixes a hexagonal Brillouin zone (BZ) for its (111) surface (Fig. 1d) on which $\bar{M}$ and $\bar{\Gamma}$ are the time-reversal invariant momenta (TRIMs) or the surface Kramers points. We carried out high-momentum-resolution angle-resolved photoemission spectroscopy (ARPES) measurements on the (111) plane of naturally grown Bi$_{2}$Se$_{3}$ (see the Methods section). The electronic spectral weight distributions observed near the $\bar{\Gamma}$ point are presented in Fig. 1a-c. Within a narrow binding-energy window, a clear V-shaped band pair is observed to approach the Fermi level (E$_{F}$). Its dispersion or intensity had no measurable time dependence within the duration of the experiment. The `V' bands cross E$_F$ at 0.09 $\AA^{-1}$ along $\bar{\Gamma}$-$\bar{M}$ and at 0.10 $\AA^{-1}$ along $\bar{\Gamma}$-$\bar{K}$, and have nearly equal band velocities, approximately $5\times 10^5$ ms$^{-1}$, along the two directions. A continuum-like manifold of states - a filled U-shaped feature - is observed inside the V-shaped band pair. All of these experimentally observed features can be identified, to first order, by a direct one-to-one comparison with the LDA band calculations. Figure 1f shows the theoretically calculated (see the Methods section)(111)-surface electronic structure of bulk Bi$_{2}$Se$_{3}$ along the $\bar{K}-\bar{\Gamma}-\bar{M}$ $k$-space cut. The calculated band structure with and without SOC are overlaid together for comparison. The bulk band projection continuum on the (111) surface is represented by the shaded areas, blue with SOC and green without SOC. In the bulk, time-reversal symmetry demands $E(\vec{k},\uparrow)=E(-\vec{k},\downarrow)$ whereas space inversion symmetry demands $E(\vec{k},\uparrow)=E(-\vec{k},\uparrow)$. Therefore, all the bulk bands are doubly degenerate. However, because space inversion symmetry is broken at the terminated surface in the experiment, SSs are generally spin-split on the surface by spin-orbit interactions except at particular high-symmetry points-the Kramers points on the surface BZ. In our calculations, the SSs (red dotted lines) are doubly degenerate only at $\bar{\Gamma}$ (Fig. 1f). This is generally true for all known spin-orbit-coupled material surfaces such as gold \cite{25,26} or Bi$_{1-x}$Sb$_x$ (ref. 5). In Bi$_{2}$Se$_{3}$, the SSs emerge from the bulk continuum, cross each other at $\bar{\Gamma}$, pass through the Fermi level ($E_{F}$) and eventually merge with the bulk conduction-band continuum, ensuring that at least one continuous band-thread traverses the bulk bandgap between a pair of Kramers points. Our calculated result shows that no surface band crosses the Fermi level if SOC is not included in the calculation, and only with the inclusion of the realistic values of SOC (based on atomic Bi) does the calculated spectrum show singly degenerate gapless surface bands that are guaranteed to cross the Fermi level. The calculated band topology with realistic SOC leads to a single ring-like surface FS, which is singly degenerate so long as the chemical potential is inside the bulk bandgap. This topology is consistent with the Z$_2=-1$ class in the Fu-Kane-Mele classification scheme \cite{7}.

A global agreement between the experimental band structure
(Fig. 1a-c) and our theoretical calculation (Fig. 1f) is obtained by considering a rigid shift of the chemical potential by about 200 meV with respect to our calculated band structure (Fig. 1f) of the formula compound Bi$_{2}$Se$_{3}$. The experimental sign of this rigid shift (the raised chemical potential) corresponds to an electron doping of the
Bi$_{2}$Se$_{3}$ insulating formula matrix (see Supplementary Information). This is consistent with the fact that naturally grown Bi$_{2}$Se$_{3}$ semiconductor used in our experiment is n-type, as independently confirmed by our transport measurements. The natural doping of this material, in fact, comes as an advantage in determining the topological class of the corresponding undoped insulator matrix, because we would like to image the SSs not only below the Fermi level but also above it, to examine the way surface bands connect to
the bulk conduction band across the gap. A unique determination
of the surface band topology of purely insulating Bi$_{1-x}$Sb$_x$ (refs 5, 6) was clarified only on doping with a foreign element, Te. In our experimental data on Bi$_{2}$Se$_{3}$, we observe a V-shaped pure SS band
to be dispersing towards E$_F$, which is in good agreement with our calculations. More remarkably, the experimental band velocities are also close to our calculated values. By comparison with calculations combined with a general set of arguments presented above, this V-shaped band is singly degenerate. Inside this `V' band, an electron-pocket-like U-shaped continuum is observed to be present near the Fermi level. This filled U-shaped broad feature is in close
correspondence to the bottom part of the calculated conduction band continuum (Fig. 1f). Considering the n-type character of the naturally occurring Bi$_{2}$Se$_{3}$ and by correspondence to our band calculation, we assign the broad feature to correspond roughly to the bottom of the conduction band.

To systematically investigate the nature of all the band features imaged in our data, we have carried out a detailed photon-energy dependence study, of which selected data sets are presented in Fig. 2a,b. A modulation of incident photon energy enables us to probe the k$_z$ dependence of the bands sampled in an ARPES study (Fig. 2c), allowing for a way to distinguish surface from bulk contributions to a particular photoemission signal \cite{5}. Our photon-energy study did not indicate a strong k$_z$ dispersion of the lowestlying energy bands on the `U', although the full continuum does have some dispersion (Fig. 2). Some variation of the quasiparticle intensity near E$_F$ is, however, observed owing to the variation of the electron-photon matrix element. In light of the k$_z$-dependence study (Fig. 2b), if the features above -0.15 eV were purely due
to the bulk, we would expect to observe dispersion as k$_z$ moved away from the $\Gamma$-point. The lack of strong dispersion yet close one-to-one correspondence to the calculated bulk band structure suggests that the inner electron pocket continuum features are probably a mixture of surface-projected conduction-band states, which also includes some band-bending effects near the surface and the full continuum of bulk conduction-band states sampled from a few layers beneath the surface. Similar behavior is also observed in the ARPES study of other semiconductors \cite{27}. In our k$_z$-dependent
study of the bands (Fig. 2b) we also observe two bands dispersing in k$_z$ that have energies below -0.3 eV (blue dotted bands), reflecting the bulk valence bands, in addition to two other non-dispersive features associated with the two sides of the pure SS Dirac bands. The red curve is measured right at the $\Gamma$-point, which suggests that the Dirac point lies inside the bulk bandgap. Taking the bottom of the `U' band as the bulk conduction-band minimum, we estimate that a bandgap of about 0.3 eV is realized in the bulk of the undoped material. Our ARPES estimated bandgap is in good agreement with the value deduced from bulk physical measurements \cite{23} and from other calculations that report the bulk band structure \cite{20,24}. This suggests that the magnitude of band bending near the surface is not larger than 0.05 eV. We note that in purely insulating Bi$_{2}$Se$_{3}$ the Fermi level should lie deep inside the bandgap and only pure surface bands will contribute to surface conduction. Therefore, in determining the topological character of the insulating Bi$_{2}$Se$_{3}$ matrix the `U' feature is not relevant.

We therefore focus on the pure SS part. The complete surface FS
map is presented in Fig. 3. Figure 3a presents electron distribution data over the entire two-dimensional (111) surface BZ. All the observed features are centered around $\bar{\Gamma}$. None of the three TRIMs located at $\bar{M}$ are enclosed by any FS, in contrast to what is observed in Bi$_{1-x}$Sb$_x$ (ref. 5). The detailed spectral behavior around $\bar{\Gamma}$ is shown in Fig. 3b, which was obtained with high momentum resolution. A ring-like feature formed by the outer `V' pure SS band (a horizontal cross-section of the upper Dirac cone in Fig. 1) surrounds the conduction-band continuum centered at $\bar{\Gamma}$. This ring is singly degenerate from its one-to-one correspondence to band calculation. An electron encircling the surface FS that encloses a TRIM or a Kramers point obtains a geometrical quantum phase (Berry phase) of $\pi$ mod $2\pi$ in its wavefunction \cite{7}. Therefore, if the chemical potential (Fermi level) lies inside the bandgap, as it should in purely insulating Bi$_{2}$Se$_{3}$, its surface must carry a global $\pi$ mod $2\pi$ Berry phase. In most spin-orbit materials, such as gold (Au[111]), it is known that the surface FS consists of two spin-orbit-split rings generated by two singly degenerate parabolic (not Dirac-like) bands that are shifted in momentum space from each other, with both enclosing the $\bar{\Gamma}$-point \cite{25,26}. The resulting FS topology leads to a 2$\pi$ or 0 Berry phase because the phases from the two rings add or cancel. This makes gold-like SSs topologically trivial despite their spin-orbit origin.

Our theoretical calculation supported by our experimental data
suggests that in insulating Bi$_{2}$Se$_{3}$ there exists a singly degenerate surface FS which encloses only one Kramers point on the surface Brillouin zone. This provides evidence that insulating Bi$_{2}$Se$_{3}$ belongs to the Z$_2=-1$ topological class in the Fu-Kane-Mele topological classification scheme for band insulators. On the basis of our ARPES data we suggest that it should be possible to obtain the fully undoped compound by chemically hole-doping the naturally occurring Bi$_{2}$Se$_{3}$, thereby shifting the chemical potential to lie inside the bulk bandgap. The surface transport of Bi$_{2}$Se$_{3}$ prepared as such would therefore be dominated by topological effects as it possesses only one Dirac fermion that carries the non-trivial Z$_2$ index. The existence of a large bulk bandgap (0.3 eV)
within which the observed Z$_2$ Dirac fermion state lies suggests the realistic possibility for the observation of topological effects even at room temperature in this material class. Because of the simplest possible topological surface spectrum realized in Bi$_{2}$Se$_{3}$, it can be
considered as the `hydrogen atom' of strong topological insulators. Its simplest topological surface spectrum would make it possible to observe and study many exotic quantum phenomena predicted in topological field theories, such as the Majorana fermions \cite{14}, magnetic monopole image \cite{9,10} or topological exciton condensates \cite{15}, by transport probes.

NOTE ADDED: \textbf{The counter-doping method to reach the fully undoped compound - the bulk-insulator state, proposed in this paper, and the spin-texture measurements have been achieved in a recent work in Nature:} [http://dx.doi.org/10.1038/nature08234 (2009)]

\section*{Methods}

\textbf{Theoretical calculations.} The theoretical band calculations were performed with the LAPW method in slab geometry using the WIEN2K package \cite{28}. The generalized
gradient approximation of Perdew, Burke and Ernzerhof \cite{29} was used to describe the exchange-correlation potential. SOC was included as a second variational step using scalar-relativistic eigenfunctions as basis after the initial calculation was
converged to self-consistency. The surface was simulated by placing a slab of 12 quintuple layers in vacuum. A grid of 21$\times$21$\times$1 points was used in the calculations,
equivalent to 48 k points in the irreducible BZ and 300 k points in the first BZ. To calculate the k$_z$ of the ARPES measurements ($k_z=(1/hbar)\sqrt{2m(E_{kin}cos^2\theta+V_0)}$), an inner potential V$_0$ of approximately 11.7 eV was used, given by a fit on the ARPES data at normal emission.

\textbf{Experimental methods.} Single crystals of Bi$_2$Se$_3$ were grown by melting stoichiometric mixtures of high-purity elemental Bi and Se in a 4-mm-inner-diameter quartz tube. The sample was cooled over a period of two days, from 850 to 650 $^{\circ}C$, and then annealed at this temperature for a week. Single crystals were obtained and could be easily cleaved from the boule. High-resolution ARPES measurements were then performed using 17-45 eV photons on beamline 12.0.1 of the Advanced Light Source at the Lawrence Berkeley National Laboratory and beamline 5-4 at the Stanford Synchrotron Radiation Laboratory. The energy and momentum resolutions were 15 meV and 1.5\% of the surface BZ respectively using a Scienta analyser. The samples were cleaved in situ between 10 and 55 K under pressures of less than 5$\times$10$^{-11}$ torr, resulting in shiny flat surfaces. The surface band quasiparticle signal is stable throughout the entire measurement duration. The counter-doping method to reach the fully undoped compound - the bulk-insulator state, proposed in this paper, and the spin-texture measurements have been achieved in a recent work in Nature: [http://dx.doi.org/10.1038/nature08234 (2009)]

\newpage

\begin{figure*}[h!]
\includegraphics[scale=1.0]{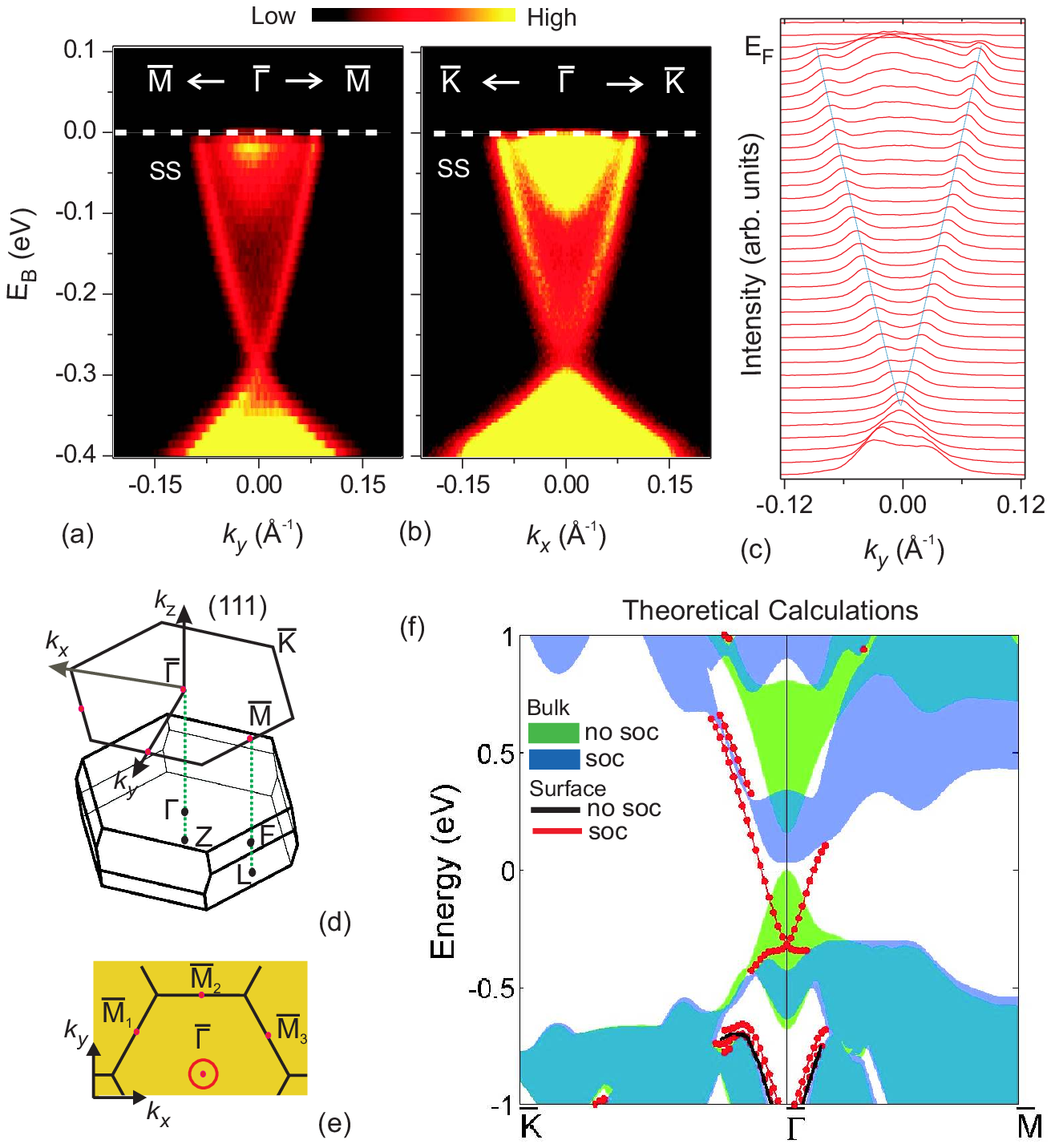}
\caption{\label{fig:charging} [Enlarged version of Fig-1] Theoretical calculations and experimental results : Strong spin-orbit interaction gives rise to a single-surface-state-Dirac-cone on the surface of the topological insulator.}
\end{figure*}
\newpage
\begin{figure*}[h!]
\includegraphics[scale=1.0]{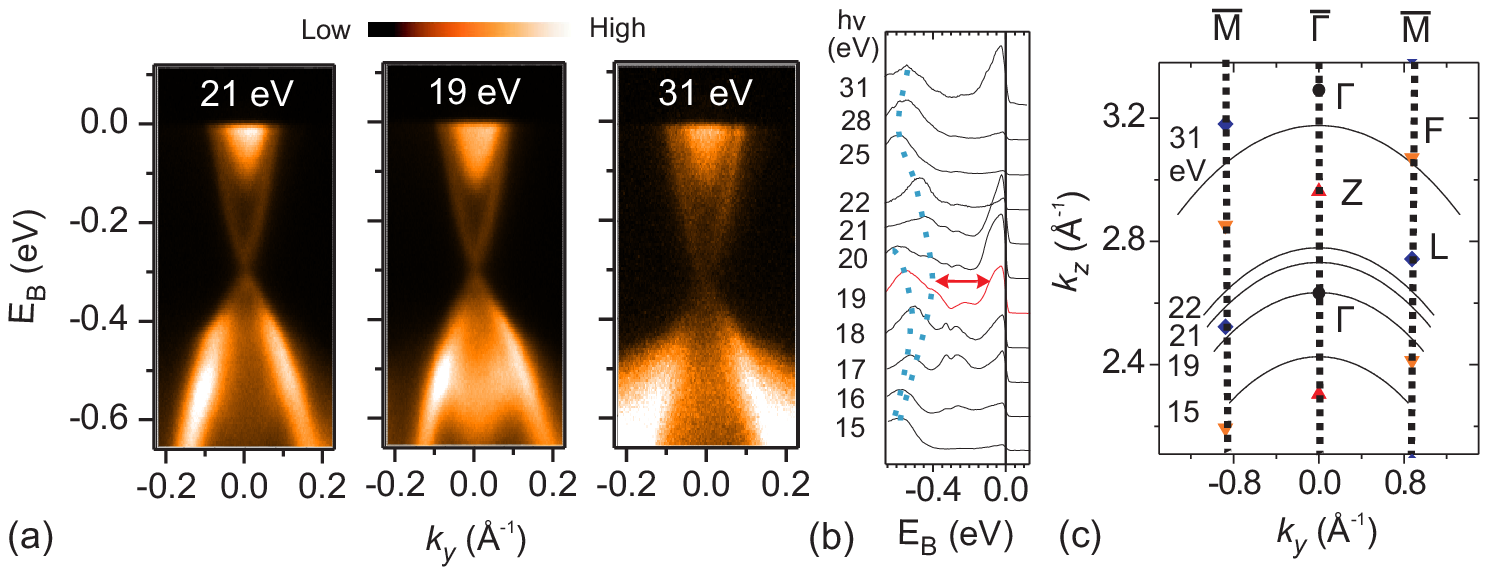}
\caption{\label{fig:charging} [Enlarged version of Fig-2] Transverse-momentum $k_z$ dependence of topological Dirac bands near $\bar{\Gamma}$.}
\end{figure*}
\newpage
\begin{figure*}[h!]
\includegraphics[scale=1.2]{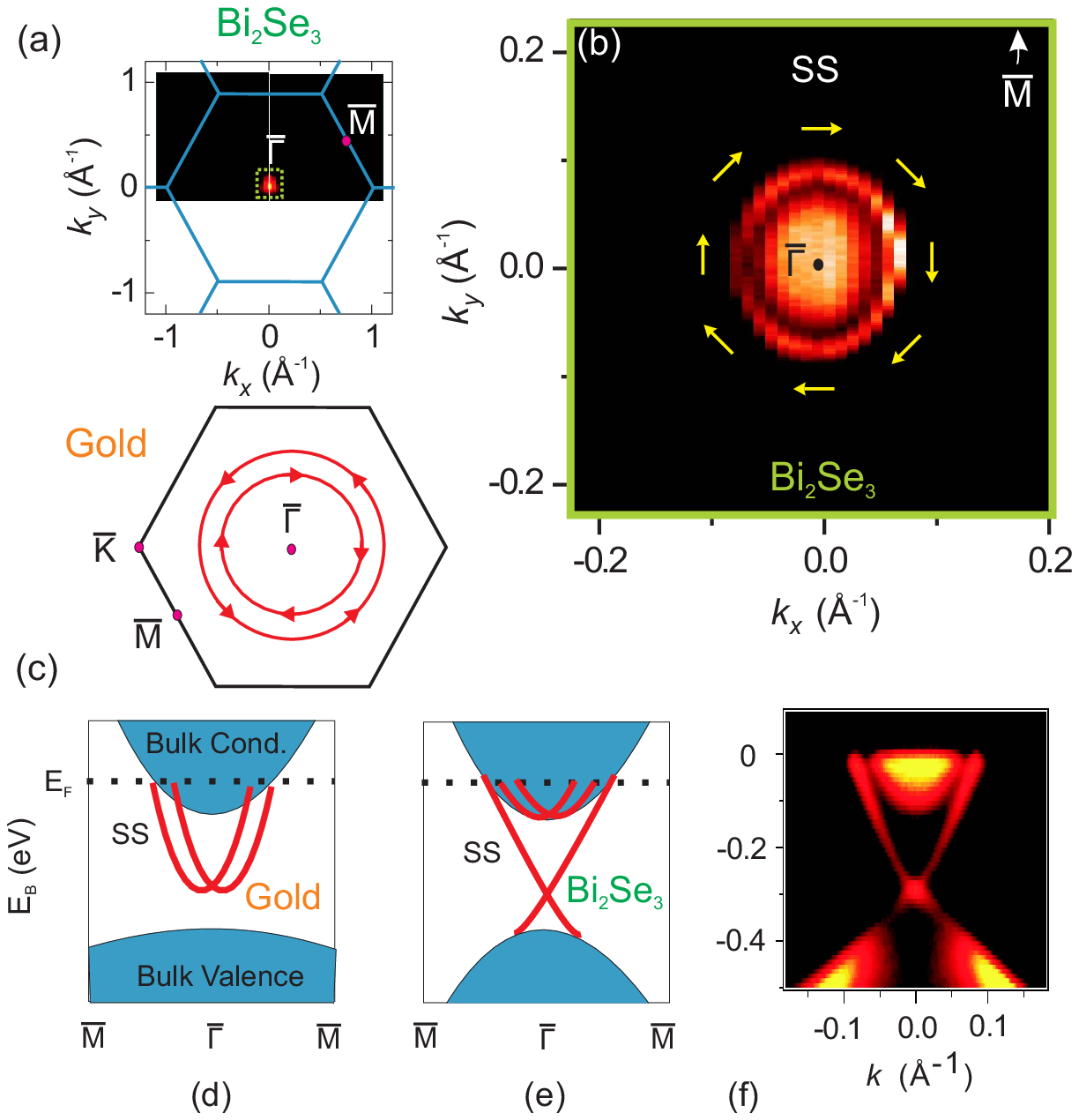}
\caption{\label{fig:charging} [Enlarged version of Fig-3] The Z$_2$ topology of the surface Dirac cone Fermi surface and the associated Berry's phase.}
\end{figure*}

\newpage
\newpage

\textbf{Fig. 1 (Caption). Strong spin-orbit interaction gives rise to a single SS Dirac cone.} Theory (see the Methods section for calculation methods) versus experiments. a,b, High-resolution ARPES measurements of surface electronic band dispersion on Bi$_2$Se$_3$ (111). Electron dispersion data measured with an incident photon energy of 22 eV near the $\bar{\Gamma}$-point along the $\bar{\Gamma}-\bar{M}$ (a) and $\bar{\Gamma}-\bar{K}$ (b) momentum-space cuts. c, The momentum distribution curves corresponding to a suggest that two surface bands converge into a single Dirac point at $\bar{\Gamma}$. The V-shaped pure SS band pair observed in a-c is nearly isotropic in the momentum plane, forming a Dirac cone in the energy-$k_x-k_y$ space (where $k_x$ and $k_y$ are in the $\bar{\Gamma}-\bar{K}$ and $\bar{\Gamma}-\bar{M}$ directions, respectively). The U-shaped broad continuum feature inside the V-shaped SS corresponds roughly to the bottom of the conduction band (see the text). d, A schematic diagram of the full bulk three-dimensional BZ of Bi$_2$Se$_3$ and the two-dimensional BZ of the projected (111) surface. e, The surface Fermi surface (FS) of the two-dimensional SSs along the $\bar{K}-\bar{\Gamma}-\bar{M}$ momentum-space cut is a single ring centered at $\bar{\Gamma}$ if the chemical potential is inside the bulk bandgap. The band responsible for this ring is singly degenerate in theory. The TRIMs on the (111) surface BZ are located at $\bar{\Gamma}$ and the three $\bar{M}$ points. The TRIMs are marked by the red dots. In the presence of strong spin–orbit coupling (SOC), the surface band crosses the Fermi level only once between two TRIMs, namely $\bar{\Gamma}$ and $\bar{M}$; this ensures the existence of a $\pi$ Berry phase on the surface. f, The corresponding local density approximation (LDA) band structure (see the Methods section). Bulk band projections are represented by the shaded areas. The band-structure topology calculated in the presence of SOC is presented in blue and that without SOC is in green. No pure surface band is observed to lie within the insulating gap in the absence of SOC (black lines) in the theoretical calculation. One pure gapless surface band is observed between $\bar{\Gamma}$ and $\bar{M}$ when SOC is included (red dotted lines).

\newpage

\textbf{Fig. 2 (Caption). Transverse-momentum $k_z$ dependence of topological Dirac bands near $\bar{\Gamma}$.} a, The energy dispersion data along the $\bar{\Gamma}$-$\bar{M}$ cut, measured with the photon energy of 21 eV (corresponding to 0.3 $k$-space length along $\Gamma$-Z $\parallel$ $k_z$), 19 eV ($\Gamma$) and 31 eV (-0.4 $k$-space length along $\Gamma$-Z of the bulk three-dimensional $\Gamma$ BZ) are shown. Although the bands below -0.4 eV binding energy show strong $k_z$ dependence, the linearly dispersive Dirac-like bands and the U-shaped
broad feature show weaker $k_z$ dispersion. The Dirac point is observed to lie inside the bulk bandgap. A careful look at the individual curves reveals some $k_z$ dependence of the U-shaped continuum (see b for details). b, The energy distribution curves obtained from the normal-emission spectra measured using
15-31 eV photon energies reveal two dispersive bulk bands below -0.3 eV (blue dotted lines). This is in addition to the two non-dispersive peaks from the Dirac-cone bands inside the gap. The Dirac band intensity is strongly modulated by the photon energy changes due to the matrix-element effects (which is
also observed in BiSb; ref. 5). c, A k-space map of locations in the bulk three-dimensional BZ scanned by the detector at different photon energies over a theta ($\theta$) range of $\pm30^{\circ}$. This map ($k_z$, $k_y$, $E_{photon}$) was used to explore the $k_z$ dependence of the observed bands.

\newpage

\textbf{Fig. 3 (Caption). The topology of the surface Dirac cone Fermi surface and the associated Berry's phase.} a, The observed surface FS of Bi$_2$Se$_3$ consists of a small electron pocket around the center of the BZ, $\bar{\Gamma}$. b, High-momentum-resolution data around $\bar{\Gamma}$ reveal a single ring formed by the pure SS V-shaped Dirac band. For the naturally occurring Bi$_2$Se$_3$, the spectral intensity in the middle of the ring is due to the presence of the 'U' feature, which roughly images the bottom of the conduction-band continuum (see the text). The observed topology of the pure surface FS of Bi$_2$Se$_3$ is different from that of most other spin-orbit materials such as gold (Au(111)). c, The Au(111) surface FS features two rings (each non-degenerate) surrounding the $\bar{\Gamma}$ point. An electron encircling the gold FS carries a Berry phase of zero,
characteristic of a trivial band insulator or metal, and can be classified by Z$_2=+1$ (ref. 7). The single surface FS observed in Bi$_2$Se$_3$ is topologically distinct from that of gold. The single non-degenerate surface FS enclosing a Kramers point ($\bar{\Gamma}$) constitutes the key signature of a topological-insulator phase characterized by Z$_2=-1$. d, e, Schematic SS topologies in gold and Bi$_2$Se$_3$ for direct comparison. In gold, the chemical potential can be continuously tuned to be placed inside the interband gap, making the SSs fully gapped. In Bi$_2$Se$_3$, however, the Dirac structure of the SSs required by the Kramers degeneracy and time-reversal invariance ensures that they remain gapless independent of the location of the chemical potential within the bulk gap. If the chemical potential is placed inside the gap, as it would naturally lie in the purely insulating undoped Bi$_2$Se$_3$, the surface transport would be dominated by the single Dirac fermion, which is of purely topological origin.

\end{document}